\documentclass [12pt]{article}

\begin{document}
\begin{center}
\textbf{\huge General Logarithmic Corrections to
Bekenstein-Hawking Entropy}
\end{center}
\begin{center}
Zhao Ren$^{a,b,}$\footnote{E-mail address:
zhaoren2969@yahoo.com.cn},
Zhao Hai-Xia$^b$ and Hu Shuang-Qi$^b$\\
$^a$Department of Physics, Shanxi Datong University, Datong 037009
P.R.China\\

$^b$School of Chemical Engineering and Environment, North
University
of China, Taiyuan 030051 P.R.China\\

\end{center}

\vspace{0.5cm}
\begin{center}
\textbf{\Large Abtract}
\end{center}

Recently, there has been a lot of attention devoted to resolving
the quantum corrections to the Bekenstein-Hawking entropy of the
black hole. In particular, the coefficient of the logarithmic term
in the black hole entropy correction has been of great interest.
In this paper, the black hole is corresponded to a canonical
ensemble in statistics by radiant spectrum, resulted from the
black hole tunneling effect studies and the partition function of
ensemble is derived. Then the entropy of the black hole is
calculated. When the first order approximation is taken into
account, the logarithmic term of entropy correction is consistent
with the result of the generalized uncertainty principle. In our
calculation, there are no uncertainty factors. The prefactor of
the logarithmic correction and the one if fluctuation is
considered are the same. Our result shows that if the thermal
capacity is negative, there is no divergent term. We provide a
general method for further discussion on quantum correction to
Bekenstein-Hawking entropy. We also offer a theoretical basis for
comparing string theory and loop quantum gravitaty and deciding
which one is
more reliable.\vspace{0.5cm}
\vspace{1.cm}\\
\textbf{\Large 1. Introduction}

One of the most remarkable achievements in gravitational physics
was the realization that black holes have temperature and entropy
[1 - 3]. There is a growing interest in the black hole entropy.
Because entropy has statistical physics meaning in the
thermodynamic system, it is related to the number of microstates
of the system. However, in the general relativity of Einstein's
theory, the black hole entropy is a pure geometrical quantity. If
we compare the black hole with the general thermodynamics system,
we can find out an important difference between them: the black
hole is a void with strong gravitation, while an ordinary object
is made up of atoms and molecules. Based on the microstructure of
general thermodynamics system, we can use the statistic mechanics
of microscopic components to explain the thermodynamic property of
an object. However, whether the black hole has the inner freedom
degree corresponding to its entropy is the key issue in the black
hole's physics  [4]. Let us suppose that the Bekenstein-Hawking
entropy can be attributed a definite statistical meaning. Then how
are these microstates defined or, in other words, how can they be
counted? [5] This is a key problem in the black hole entropy. In
recent years, string theory and loop quantum gravitaty both have
been successful in statistically explaining the black hole
entropy-area law [5]. Which one is more reliable? It is expected
to make choice by discussing the quantum correction term of the
black hole entropy. Therefore, studying the black hole entropy
correction value becomes the focus of attention. Many ways of
discussing the black hole entropy correction value have emerged
[5-12]. (But over the last few years both are String Theory and in
loop quantum gravity.)

Based on string theory and loop quantum gravitaty, the
relationship of the black hole entropy-area is given by[13]

\begin{equation}
\label{eq1}
S = \frac{A}{4L_p^2 } + \rho \ln \frac{A}{4L_p^2 } + O\left( {\frac{L_p^2
}{A}} \right),
\end{equation}

\noindent where $A = 16\pi L_p^2 M^2$ is area of the black hole
horizon, $L_p = \sqrt {\hbar G} $ is Planck length. For the case
of Loop Quantum Gravity, which is here of interest, there is still
no consensus on the coefficient of the logarithmic correction,
$\rho $, but it is established [14-16] that there are no
correction terms with stronger-than-logarithmic dependence on the
area.

Most of the recent focus has been on $\rho $; that is, the
coefficient of the leading-order correction or the logarithmic
``prefactor''. It has even been suggested that this particular
parameter might be useful as a discriminator of prospective
fundamental theories [16]. It is, therefore, appropriate to
reflect upon the loop quantum gravity prediction of $\rho = - 1 /
2$ (according to the most up-to-date rigorous calculation [17]);
whereas string theory makes no similar type of assertion [5].
However, without any further input, how can we determine if any
particular value of the prefactor is right or wrong? That is,
unlike the tree-level calculations, this type of discrimination is
based on a question to which we do not know the answer yet!

It becomes clear that, to proceed in this direction, one requires
a method of fixing $\rho $ indepent of the specific elements of
any one particular model of quantum gravity. Recently Hawking
radiation process of the black hole is given a new
explanation-tunneling process. And the radiation spectrum of the
black hole is obtained. Thereafter, using this radiant spectrum,
we calculate the partition function of canonical ensemble and
derive the correction term of the black hole entropy. In our
calculation, there is no need to make any hypothesis. We provide a
new method for the correction to
the black hole entropy studies.\vspace{1.cm}\\
\textbf{\Large 2. Calculation and analysis}

Recently, Parikh, Wilczek and Kraus [18] discussed Hawking
radiation by tunneling effect. They thought that tunnels in the
process of the particle radiation of the black hole has no
potential barrier before particles radiate. Potential barrier is
produced by radiation particles itself. That is, during the
process of tunneling effect creation, the energy of the black hole
decrease and the radius of the black hole horizon reduces. The
horizon radius becomes a new value that is smaller than the
original value. The decrease of radius is determined by the value
of energy of radiation particles. There is a classical forbidden
band-potential barrier between original radius and the one after
the black hole radiates. Parikh and Wilczek skillfully obtained
the radiation spectrum of Schwarzschild and Reissner-Nordstrom
black holes. Refs.[19-30] developed the method proposed by Parikh
and Wilczek. They derived the radiation spectrum of the black hole
in all kinds of space-time. Refs.[31-34] obtained radiation
spectrum of Hawking radiation after considering the generalized
uncertainty relation. However, Angheben, Nadalini, Vanzo and
Zerbini calculated the radiation spectrum of arbitrary dimensional
black hole and obtained that the general expression of the energy
spectrum of Hawking radiation is as follows:

\begin{equation}
\label{eq2}
\rho _s \propto e^{\Delta S},
\end{equation}

\noindent
where

\[
\Delta S = S_{MC} (E - E_s ) - S_{Mc} (E)
 = \sum\limits_{k = 1} {\frac{1}{k!}\left( {\frac{\partial ^kS_{MC} (E_b
)}{\partial E_b^k }} \right)_{E_s = 0} } ( - E_s )^k
\]

\begin{equation}
\label{eq3}
 = - \beta E_s + \beta _2 E_s^2 + \cdots ,
\end{equation}
$E_b = E - E_s ,$ according to thermodynamics relation, $\beta $
should be the inverse of the temperature,

\begin{equation}
\label{eq4} \beta _k = \frac{1}{k!}\left( {\frac{\partial ^k\ln
\Omega }{\partial E_b^k }} \right)_{_s = 0}
 = \frac{1}{k!}\left( {\frac{\partial ^k S_{MC} }{\partial E_b^k }}
\right)_{E_s = 0} .
\end{equation}
Normalizing  $\rho _s $ in (2), we obtain

\[
\rho _s = \frac{1}{Z_c}e^{S_{MC} (E - E_s ) - S_{MC} (E)},
\]

\noindent where $Z_c$ is called canonical partition function: $Z_c
= \sum\limits_s {e^{S_{MC} (E - E_s ) - S_{MC} (E)}} $ We begin
with the formula for the canonical partition function of a
classical system in equilibrium

\begin{equation}
\label{eq5} Z_c (\beta ) = \int\limits_0^\infty {e^{\Delta S}dE_s}
\rho (E_b),
\end{equation}

\noindent where, $\rho (E_b)$ is the density of states. In what
follows, we shall employ the identification $\rho (E_b) \equiv
e^{S_{MC} (E_b)}$ [35], where, $S_{MC} (E_b)$ is the
microcanonical entropy of an isolated subsystem whose energy is
fixed at $E_b$. According to (\ref{eq3}), the energy of the black
hole radiant particles is $E_s $, then the energy of the black
hole is $E_b = E - E_s $, for the black hole, the state density
with energy $E_b $ is $\rho (E - E_s )$. So

\begin{equation}
\label{eq6}
\rho (E - E_s ) = \exp \left[ {S_{MC} (E - E_s )} \right].
\end{equation}
The integral in Eq (\ref{eq5}) can be performed in general by the
saddle point approximation, provided the microcanonical entropy
$S_{MC} (E - E_s )$ can be Taylor-expanded around the average
equilibrium energy $E$,

\begin{equation}
\label{eq7}
S_{MC} (E - E_s ) = S_{MC} (E) - \beta E_s + \beta _2 E_s^2 + \cdots ,
\end{equation}

\noindent and higher order terms in powers of the energy
fluctuation represented by the $\ldots $ in this expansion can
neglected in comparison to the terms of second order.

(\ref{eq5}) can be rewritten as

\[
Z_c (\beta ) = \int\limits_0^\infty {e^{ - \beta E_s + \beta _2 E_s^2 }dE_s
} e^{S_{MC} (E - E_s )}
 = e^{S_{_{MC} } (E)}\int\limits_0^\infty dE_s{e^{ - 2\beta E_s + 2\beta _2
E_S^2 }}
\]

\begin{equation}
\label{eq8}
 = e^{S_{MC} (E)}
\left[ {\frac{1}{2}\sqrt {\frac{\pi }{ - 2\beta _2 }} \exp \left(
{\frac{\beta ^2}{ - 2\beta _2 }} \right)\left( {1 - erf\left( {\frac{\beta
}{\sqrt { - 2\beta _2 } }} \right)} \right)} \right].
\end{equation}

\noindent
where

\[
erf(z) = \frac{2}{\sqrt \pi }\int\limits_0^z {e^{ - t^2}} dt
\]

\noindent
is error integral.

Using the standard formula from equilibrium statistical mechanics

\begin{equation}
\label{eq9} S = \ln Z_c - \beta \frac{\partial \ln Z_c}{\partial
\beta },
\end{equation}

\noindent
it is easy to deduce that the canonical entropy is given in terms of the
microcanonical entropy by

\begin{equation}
\label{eq10}
S_C (E) = S_{MC} (E) + \Delta _S ,
\end{equation}

\noindent
where

\begin{equation}
\label{eq11}
\Delta _S = \ln f(\beta ) - \beta \frac{\partial \ln f(\beta )}{\partial
\beta },
\end{equation}

\begin{equation}
\label{eq12}
f(\beta ) = \frac{1}{2}\sqrt {\frac{\pi }{ - 2\beta _2 }} \exp \left(
{\frac{\beta ^2}{ - 2\beta _2 }} \right)
\left[ {1 - erf\left( {\frac{\beta }{\sqrt { - 2\beta _2 } }} \right)}
\right].
\end{equation}
According to the asymptotic expression of error function

\[
erf(z) = 1 - \frac{e^{ - z^2}}{\sqrt \pi z}\left[ {1 + \sum\limits_{k =
1}^\infty {( - 1)^k\frac{(2k - 1)!!}{(2z^2)^k}} } \right],
\quad
\left| z \right| \to \infty ,
\]
We have

\begin{equation}
\label{eq13}
f(\beta ) = \frac{1}{2\beta }\left[ {1 + \sum\limits_{k = 1}^\infty {( -
1)^k\frac{(2k - 1)!!}{2^k}\left( {\frac{\sqrt { - 2\beta _2 } }{\beta }}
\right)} ^{2k}} \right].
\end{equation}
Substituting (\ref{eq13}) into (\ref{eq11}), we derive

\[
\Delta _S = \ln \left[ {\frac{1}{2\beta } + \sum\limits_{k = 1}^\infty {( -
1)^k\frac{(2k - 1)!!}{2^k2\beta }\left( {\frac{\sqrt { - 2\beta _2 } }{\beta
}} \right)^{2k}} } \right]
\]

\begin{equation}
\label{eq14}
 + \frac{1 + \sum\limits_{k = 1}^\infty {( - 1)^k(2\sqrt { - 2\beta _2 }
)^{2k}\frac{(2k + 1)(2k - 1)!!}{2^k(2\beta )^{2k}}} }{1 + \sum\limits_{k =
1}^\infty {( - 1)^k(2\sqrt { - 2\beta _2 } )^{2k}\frac{(2k -
1)!!}{2^k(2\beta )^{2k}}} }.
\end{equation}
Using the definition of the thermal capacity of the system

\begin{equation}
\label{eq15}
C \equiv - \beta ^2\left( {\frac{\partial E}{\partial \beta }} \right),
\end{equation}

\noindent
and

\begin{equation}
\label{eq16}
\beta _2 = - \frac{1}{2}\frac{\beta ^2}{C}.
\end{equation}
(\ref{eq14}) can be expressed as

\begin{equation}
\label{eq17} \Delta _S = \ln \left[ {T + T\sum\limits_{k =
1}^\infty {( - 1)^k\frac{(2k - 1)!!}{2^kC^k}} } \right]
 + \frac{1 + \sum\limits_{k = 1}^\infty {( - 1)^k\frac{(2k + 1)(2k -
1)!!}{2^kC^k}} }{1 + \sum\limits_{k = 1}^\infty {( - 1)^k\frac{(2k
- 1)!!}{2^kC^k}} }.
\end{equation}

\noindent
where $T$ is the temperature of the system. When we take the first order
approximation, the logarithmic correction term of entropy is

\begin{equation}
\label{eq18} \Delta _S = \ln T.
\end{equation}
The correction to entropy is not related to thermal capacity.

For error function, if we take the sum of the series from the
first term to $n$ the term as the approximation of $erf(z)$. If
$z$ is real number, radical error does not exceed the absolute
value of the first term in the series. So if $C < -1$ or $C > 1$,
it is
sure that $\Delta _S $ is not divergent.\vspace{1.cm}\\
\textbf{\Large 3. Conclusion }

In summary, for Schwarzschild space-time, when we only take the
first order approximation, the logarithmic correction term of
entropy is

\begin{equation}
\label{eq19} \Delta _S = \ln T = - \frac{1}{2}\ln \frac{A}{4} +
const.
\end{equation}
Using generalized uncertainty principle, Ref.[5] obtained that
correction term of the black hole entropy was

\begin{equation}
\label{eq20} S
 = \frac{A}{4} - \frac{\pi \alpha ^2}{4}\ln \left( {\frac{A}{4}} \right) +
\sum\limits_{n = 1}^\infty {C_n } \left( {\frac{A}{4}} \right)^{ - n} +
const.
\end{equation}
From (\ref{eq20}), we know that the logarithmic term of the black
hole entropy correction contains uncertainty factor $\alpha ^2$.
However, the uncertainty does not appear in our result.

After considering the correction to the black hole thermodynamics
quantities due to thermal fluctuation, the expression of the
entropy is [36-39]

\begin{equation}
\label{eq21} S = \ln \rho = S_{MC} - \frac{1}{2}\ln (CT^2) +
\cdots ,
\end{equation}
The above mentioned result is evidently limited. The thermal
capacity of Schwarzschild black hole is negative causing the
logarithmic correction term of the entropy in (\ref{eq21}) to be
divergent. So this relation is not valid for Schwarzschild black
hole. However, any general four-dimensional curved space-time can
be reduced to Schwarzschild space-time with proper approximation
or limit. Hence (\ref{eq21}) does not represent general property.
However, in our result, the condition $C < -1$ or $C > 1$ can be
satisfied by almost all black holes. So it is general.

In addition, previous results of the black hole entropy by other
researchers were based on the fact that the black hole has thermal
radiation and the radiation spectrum is a pure thermal one.
However, Hawking radiation derives pure thermal spectrum under the
condition that the space-time is invariant. The dispute over the
loss of information arises while the radiation is dealt with. The
black hole information loss means that pure quantum state will
disintegrate to mixed state. This violates the unitary positive
principle in quantum mechanics. Applying the tunneling effect
method, we study the black hole radiation. Considering energy
conservation and the change of horizon, the radiation spectrum is
no longer a strictly pure thermal spectrum. This method avoids the
limitation of Hawking radiation. It is pointed out that the
self-gravity action provides the potential barrier of the quantum
tunnel.

Our discussion is based on the quantum tunneling effect of the
black hole radiation. Our result is very reasonable. We provide a
new method for further studing the quantum correction to
Bekenstein-Hawking entropy. We also offer a theoretical basis for
comparing the string theory and loop quantum gravitaty and
deciding which one is more reliable? Our result has shown that if
the thermal capacity satisfies $-1 \le C \le 1$, the logarithmic
correction term of the black hole entropy may be divergent. What
does this divergent mean? This problem needs further
investigation.\\
\textbf{ACKNOWLEDGMENT}

This project was supported by the  Shanxi
Natural Science Foundation of China under Grant No. 2006011012 \\
\textbf{References}

[1] J. D.Bekenstein.   Phys. Rev. \textbf{D 7}, 2333(1973)

[2] J. D.Bekenstein.   Phys. Rev. \textbf{D 9}, 3292(1974)

[3] S. W. Hawking.   Nature \textbf{248} 30,(1974);  Commun. Math.
Phys. \textbf{43}, 199(1975)

[4] Y. J. Wang.  Black Hole Physics (Changsha: Hunan Normal
University Press) p263(2000) (in Chinese)

[5] A. J. M. Medved and E. C. Vagenas.  Phys. Rev. \textbf{D 70},
124021(2004)

[6] A. Chatterjee and  P. Majumdar. Phys. Rev. Lett. \textbf{92},
141301(2004)

[7] M. Cavaglia, S. Das, R. Maartens,  Class. Quant. Grav.
\textbf{21}, 4511 (2004)

[8] G. A. Camellia,  M. Arzano and A. Procaccini.  Phys. Rev.
\textbf{D 70}, 107501(2004)

[9] A. Chatterjee  and P. Majumdar. Phys. Rev. \textbf{D 71},
024003 (2005)

[10] Y. S. Myung. Phys. Lett. \textbf{B 579}, 205(2004)

[11] M. M. Akbar and S. Das. Class. Quant. Grav. \textbf{21}, 1383
(2004)

[12] S. Das.  Class. Quant. Grav. \textbf{19}, 2355(2002)

[13] G. A. Camelia, M. Arzano and A. Procaccini, Phys. Rev.
\textbf{D 70}, 107501(2004)

[14] C. Rovelli, Phys. Rev. Lett. \textbf{77}, 3288(1996)

[15] A. Ashtekar, Phys. Rev. Lett. \textbf{80},  904(1998)

[16] R. K. Kaul and P. Majumdar, Phys. Rev. Lett. \textbf{84},
5255(2000)

[17] K. A. Meisser, Black hole entropy from Loop Quantum Gravity.
gr-qc/0407052

[18] M. K. Parikh and F. Wilczek, Phys. Rev. Lett. \textbf{85},
5042(2000); Esko Keski-Vakkuri, P. Kraus,  Nucl. Phys. \textbf{B
491}, 249(1997); P. Kraus and F. Wilczek,  Nucl. Phys. \textbf{B
437}, 231(1995).

[19] E. C. Vagenas, Phys. Lett. \textbf{B 503}, 399(2001)

[20] E. C. Vagenas, Mod. Phys. Lett. \textbf{A 17}, 609(2002)

[21] E. C. Vagenas, Phys. Lett. \textbf{B 533}, 302(2002)

[22] A. J. Medved, Class. Quant. Grav. \textbf{19}, 589(2002)

[23] M. K. Parikh, Phys. Lett. \textbf{B 546}, 189(2002)

[24] A. J. M. Medved, Phys. Rev. \textbf{D 66}, 124009(2002)

[25] E. C. Vagenas, Phys. Lett. \textbf{B 559}, 65(2003)

[26] J. Y. Zhang and Z. Zhao, Phys. Lett. \textbf{B 618},
14(2005); C. Z. Liu, J. Y. Zhang and Z. Zhao, Phys. Lett.
\textbf{B 639}, 670(2006)

[27] J. Y. Zhang and Z. Zhao, Nucl. Phys. \textbf{B 725},
173(2005); \textbf{JHEP10}, 055(2005)

[28] M. Angheben, M. Nadalini, L. Vanzo and S. Zerbini,
\textbf{JHEP05}, 014(2005)

[29] R. Kerner and R. B. Mann, Phys. Rev. \textbf{D 73},
104010(2006)

[30] Q. Q. Jiang and S. Q. Wu, Phys. Lett. \textbf{B 635},
151(2006)

[31] M. Arzano, A. J. M. Medved and E. C. Vagenas,
\textbf{JHEP09}, 037(2005)

[32] A. J. M. Medved and E.C. Vagenas, Mod. Phys. Lett. \textbf{A
20}, 2449(2005)

[33] A. J. M. Medved and E. C. Vagenas, Mod. Phys. Lett. \textbf{A
20}, 1723(2005)

[34] M. Arzano, Mod. Phys. Lett. \textbf{A 21}, 41(2006)

[35] A. Chatterjee  and P. Majumdar. Black hole entropy: quantum
vs thermal fluctuations. gr-qc/0303030


[36] M. R. Setare, Phys. Lett. \textbf{B 573}, 173(2003).

[37] M. Cavaglia and A. Fabbri, Phys. Rev. \textbf{D 65},
044012(2002)

[38] G. Gour, A. J. M. Medved, Class. Quant. Grav. \textbf{20},
3307(2003)

[39] M. R. Setare, Eur. Phys. J. \textbf{C 33}, 555(2004).

\end{document}